# Laser-stimulated photodetachment of electrons from the negatively charged dielectric substrates

Y. Ussenov[1], M. N. Shneider[1], S. Yatom[2], and Y. Raitses[2]

[1] *Department of Mechanical and Aerospace Engineering, Princeton University, Princeton, USA*

[2] *Princeton Plasma Physics Laboratory, Princeton University, Princeton, USA*

**Abstract**

It is shown experimentally that the photodetachment yield of surplus electrons created by plasma-induced charging of non-conductive surfaces of dielectric materials depends on the initial surface charge density and do not correlate with the tabulated affinity values of these materials. This unexpected result obtained using laser stimulated photodetachment for fused silica, boron nitride, and alumina, is critically important for understanding of charging and discharging dynamics, secondary electron emission and photo emission effects affecting plasma-wall interactions relevant to surface and capacitively coupled discharges, dusty plasmas, electrostatic probe diagnostics and applications for plasma processing of materials, plasma propulsion and gas breakdown.

The charging and discharging dynamics of non-conductive materials by plasmas is involved in plasma-surface interactions[1] and related interface phenomena[2]. This phenomenon is relevant to widespread applications of various semiconducting and insulating materials as plasma-confined walls[3], substrates for materials processing (e.g., silicon wafers)[4,5,6,7] and electrically insulating holders for diagnostic equipment such as for example, electrostatic probes[8]. Additionally, it plays a crucial role in plasma discharges supported or affected by secondary electron emission (SEE) from dielectric materials such as dielectric barrier[9] discharge (DBD) and capacitively coupled discharges, and plasmas for space applications[10,11].

The SEE from the surfaces of plasma-facing materials, stimulated by plasma species such as ions, electrons[12,13] and photons[14], constitutes a key process that defines discharge breakdown and sustainment, sheath formation[15,16], and overall ionization balance[17]. While the SEE induced by ions and electrons has been extensively studied[18,19], the role of electron emission stimulated by photons has not been thoroughly considered. However, the role of material properties in detachment of surplus electrons from dielectric materials, including but not limited to photon-induced secondary electron emission, is overlooked. For example, the energy threshold for the detachment of surplus electrons from dielectric materials may affect fundamental plasma processing such as recombination of positive ions or formation of negative ions on dielectric surfaces.

The laser-stimulated photodetachment (LSPD) is a technique frequently employed for studies of negative ions in plasmas and ion beams[20] and can also be applied for the planar surfaces. In Ref.[21], it was shown that the photodetachment from the glass ($SiO_2$) electrode surface of DBD induced a transition of the discharge from glow to the Townsend mode. The threshold binding energy of the surface charge electrons was assumed to be in the range of ~ 1.17-2.33 eV, while the electron affinity of the $SiO_2$ ~ 0.9-1.3 eV. The photodetachment yield was estimated to be ~$10^{-8}$. However, the actual values of yield and cross sections were not determined because of uncertainties in the laser beam shape. In another study, the impact of photodetachment of surplus negative charges from alumina surface by discharge-generated photons predicted to be the trigger for the streamer formation from opposite electrodes of the DBD[22]. The binding energy is assumed to be lower than 3.5 eV, while theoretical estimations[23] and thermal desorption studies from the alumina lead in prediction of the binding energy of electrons as ~1 eV[24,25]. The LSPD was also applied for diagnostics of particle charge[26,27,28]. In Ref.[27], the effect of the thermal emission found to be negligible on the detachment of electrons, however the role of photoemission and photodetachment is not clearly defined as the photon energy was close to the work function of common dielectric materials.

In all relevant previous studies, the clear distinction between photoemission and photodetachment in the visible wavelength range, when photon energy is less than the work function and their role in SEE from plasma-charged dielectric surfaces was not explored in detail. The "binding" energy of surplus electrons on plasma-charged surfaces are uncertain, and the actual photodetachment yield and cross-sections are not well-defined even for common dielectric surfaces. Commonly accepted hypotheses relying on the assumption of "tabulated" surface electron affinity of clean materials as the threshold value needs to be tested for real dielectric materials widely used in gas discharge plasma applications[29,30]. In this letter, the LSPD of electrons from plasma-charged dielectric substrates such as alumina ($Al_2O_3$), fused silica ($SiO_2$), and hexagonal boron nitride (h-BN) has been studied. Laser pulses with photon energies of 2.33 eV and 3.49 eV were applied to minimize bulk electron photoemission. The obtained photodetachment yield and cross sections have been analyzed in relation to the tabulated electron affinities $\chi$ of used dielectric samples (Table 1).

Table 1. Tabulated electron affinities and relative permittivity of studied dielectric materials

| Material | Rel. permittivity | Diameter, mm | Thickness, mm | $\chi$, eV |
|---|---|---|---|---|
| $SiO_2$ | 3.8 | 25.4 | 1.56 | 0.9-1.3[30,31] |
| $Al_2O_3$ | 9.5 | 25.4 | 1.56 | 1.9-2.5[30,31] |
| h-BN | 4.0 | 25.4 | 1.90 | 1.14 - 1.34[32] |

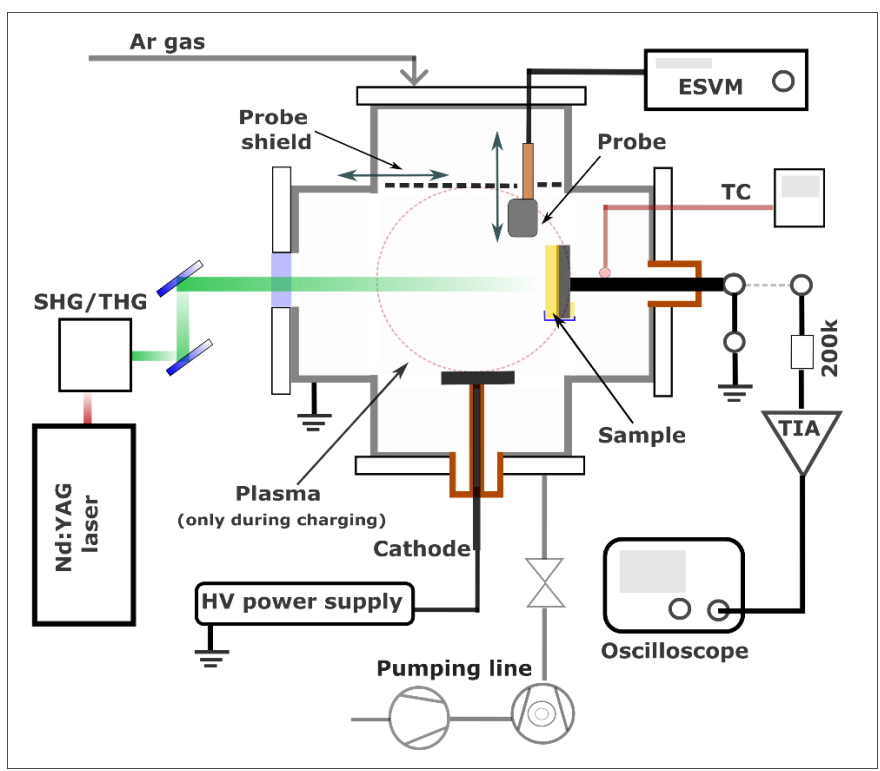

Figure 1. The general scheme of the experimental setup

The experimental setup is shown in Figure 1. The industrial-grade $SiO_2$ (Momentive Tech., 99.9%), $Al_2O_3$ (Coors Tek Inc., 99.5%) and h-BN (Momentive Tech., 95%) samples, are fixed on the metallic substrate holder isolated from the chamber walls. The samples cleaned in acetone, ethyl alcohol and distilled water to remove the surface contaminants. To eliminate the physiosorbed water from the surface, the samples heated up to ~180-200°C under the ~5x10$^{-7}$ Torr pressure for several hours. Preliminary laser beam exposures of samples were conducted at high laser fluences under the vacuum to remove the rest of the impurities from the surface and ensure good reproducibility of measurements.

The dielectric samples charged by DC glow discharge ($V_{dis}$ = -450 V, $I_{dis}$ = 1.5 ± 0.3 mA) in 100 mTorr Argon (99.999%) plasma. The stainless-steel cathode (D=25.4 mm) of the DC discharge was inserted from the opposite side to the electrostatic voltmeter (ESVM) probe and fixed at a distance of ~ 30 mm from the sample edge, while the grounded chamber wall served as the anode of the discharge.

For the LSPD, the second (SH) and third (TH) harmonics of the Nd:YAG laser (beam diameter ~ 9 mm, pulse frequency 10 Hz, and a pulse width is ~9 ns) with wavelengths of 532 nm ($hv$ = 2.33 eV) and 355 nm ($hv$ = 3.49 eV) are exposed perpendicular to the sample surface through a glass viewport. The change in the surface potential measured with Trek –347 non-contact ESVM probe positioned perpendicular to the substrate surface at 1.0±0.3 mm distance. During plasma charging, the ESVM probe head is retracted, and a metallic shield plate is placed between the probe and the discharge region. The photodetached electron current from the samples measured by the home made transimpedance amplifier (TIA) for the 355 nm laser wavelength. The current and surface voltage by ESVM were measured after separate shots to compare the photodetachment yield obtained from surface potential and photodetached electron current measurements. The TIA connected to the substrate holder via a 200 kΩ resistor and current signals are recorded by an oscilloscope (Siglent -SDS2354, 350 MHz, 2 Gs/s), with the output synchronization signal from the laser control unit used as the trigger.

Figure 2 shows the surface potential decay for the dielectric samples during the continuous laser exposure with 532 nm and 355 nm wavelength and 0.030 J/cm² and 0.019 J/cm² beam fluence respectively. After plasma charging the surface potential of all samples was $V_{surf} = -31.0 \pm 4\ V$ that is due to charging by plasma electrons. The negative surface potential decay rate with 532 nm laser exposure is fast for the first ~10³ of laser shots leading to the reduction (in absolute value) of the surface potential down to $V_{surf} \sim -17.5\ V$ for Al$_2$O$_3$ and h-BN and $V_{surf} \sim -28.0\ V$ for SiO$_2$. Further decay of the surface potential is very slow and after 9.6 x 10³ pulses it drops no more than $\Delta V_{surf} = 10V$. For the 355 nm laser exposure, the surface potential decay is faster depending on the number of the laser shots. Note that the discharging time scale for the 355 nm laser in Figure 3 (c) and (d) is an order of magnitude less than that for the 532 nm laser. The first ~200 shots cause the potential drop of about $\Delta V_{surf} = 24V$ for BN and Al$_2$O$_3$, and $\Delta V_{surf} = 20V$ for SiO$_2$. The corresponding surface electron charge density decay calculated according to the expression $\sigma_e = \Delta V_{surf} \varepsilon \varepsilon_0 / d$, where $\varepsilon_0$ – electric constant, $\varepsilon$ –dielectric permittivity and $d$ – thickness of substrate. Note that the charge density for SiO$_2$ and h-BN shows the similar value $\sigma_e \approx -0.6 \; \mu C/m^2$, while for the Al$_2$O$_3$ $\sigma_e \approx -1.65 \; \mu C/m^2$ which is almost ~3 times higher due to the large dielectric permittivity (Table 1) at the similar thickness and surface area.

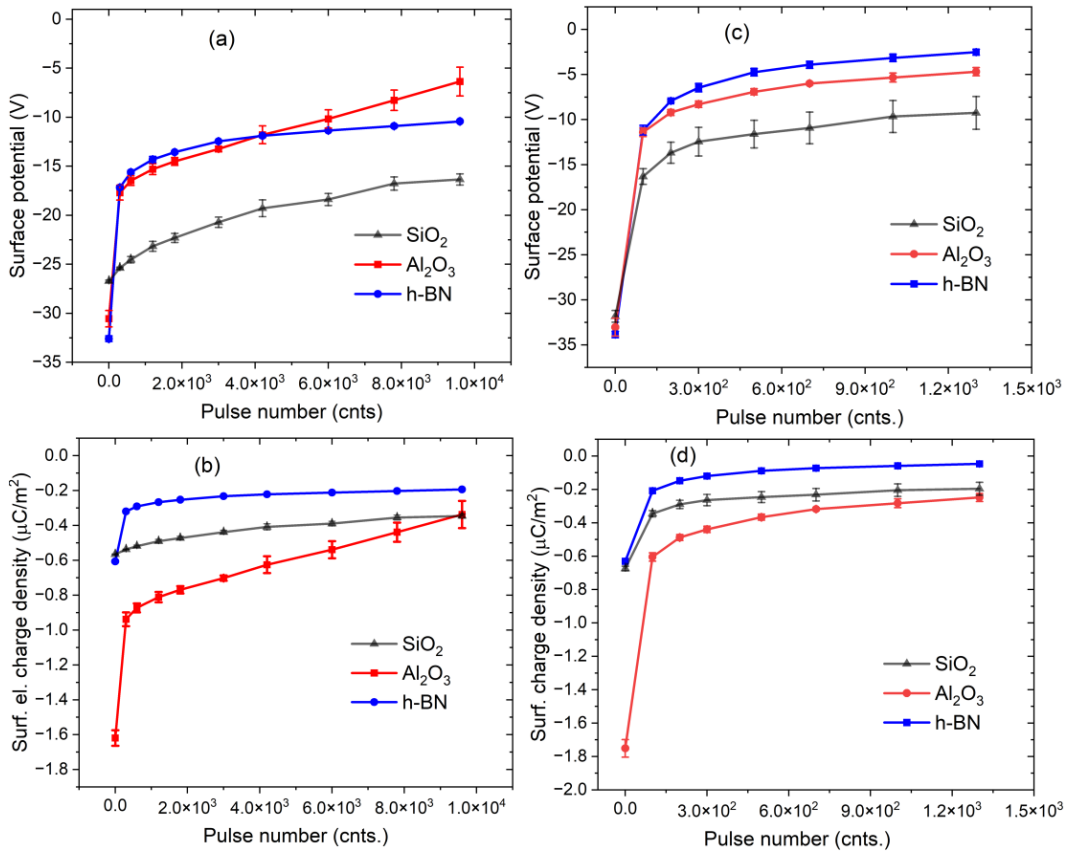

Figure 2. The surface potential and charge density decay for the dielectric samples by LSPD with 532 nm, 0.030 J/cm$^2$ (a,b) and 355 nm, 0.019 J/cm$^2$ (c,d) wavelength laser beams.

Figure 2 demonstrates that the LSPD efficiency and the rate of charge decay is not linear depending on laser shots. This behavior indicates that the LSPD yield is not constant and falls after each laser shot. The average yield per pulse is calculated as:

$$\gamma_{pd} = \frac{\sigma_e}{\sigma_{ph}} = \frac{\Delta V \, \varepsilon \varepsilon_0 \, S_{las}}{d \, e \, N_{pulse} \, n_{ph.p}}. \tag{1}$$

Here, $\sigma_e = \frac{\Delta Q}{S_s \, N_{pulse}} = (\Delta V_{surf} \, \varepsilon \varepsilon_0)/(d \, e \, N_{pulse})$ is the density of detached electrons per unit area, where $\Delta Q = C \Delta V_{surf}$ is the amount of detached charge at certain number of laser shots, $C = \varepsilon \varepsilon_0 S_s / d$ is the substrate capacitance, $S_s$ is the substrate surface area, $N_{pulse}$ is the number of laser shots, $\sigma_{ph} = n_{ph.p}/S_{las}$ is the surface density of incident photons, $n_{ph.p}$ is the number of photons per pulse and $S_{las}$ is the laser beam cross section area.

Examples of the average LSPD yield for different substrate materials and laser photon energy listed in Table 2. The highest photodetachment yield shows the Al₂O₃, then h-BN, and LSPD yield for SiO₂ shows minimum value.

Table 2. The average LSPD yield of samples at different laser photon energy.

| Material | 355 nm (3.49 eV) | 532 nm (2.33 eV) |
|---|---|---|
| Al$_2$O$_3$ | 2.06±0.67 x 10$^{-10}$ | 1.75±0.35 x 10$^{-11}$ |
| h-BN | 7.46±2.46 x 10$^{-11}$ | 7.36±1.47 x 10$^{-12}$ |
| SiO$_2$ | 5.83±1.92 x 10$^{-11}$ | 7.27±1.45 x 10$^{-13}$ |

To validate the obtained LSPD yields from the ESVM measurements, the separate experiments done with the 355 nm laser shots and based on the detecting by TIA the current pulses induced by the LSPD. Figure 3(a) shows the photodetachment current for the first pulse measured for different samples. It was found that the number of detached electrons reduces for subsequent laser shots for the same laser fluence and confirms the results of potential decay and yield from the surface potential measurements with ESVM. Figure 3(b) shows the photodetachment currents normalized to the maximum value, which show the similar decay time constant of ~5 μs for all the substrate samples. This demonstrates that the decay time is not defined by the capacitance of the dielectric substrates, but by the capacitance of the vacuum chamber, which is the same for all the case. The detachment yield is defined as:

$$\gamma_{pd} = \frac{Q_{ph.det.}}{n_{ph.p}}, \tag{2}$$

where $Q_{ph.det.}$ -number of detached electrons in elementary charge units and calculated by $Q_{ph.det.} = \int_{t_0}^{t_1} I(t)dt$, where $I(t)$ -measured photodetachment current.

Table 3 shows a comparison of the results for the first pulse for all three dielectric samples. The results from both measurements are in reasonable agreement within the errors and demonstrate that the $Al_2O_3$ has highest LSPD yield compared to the other samples, and $SiO_2$ showed the lowest value.

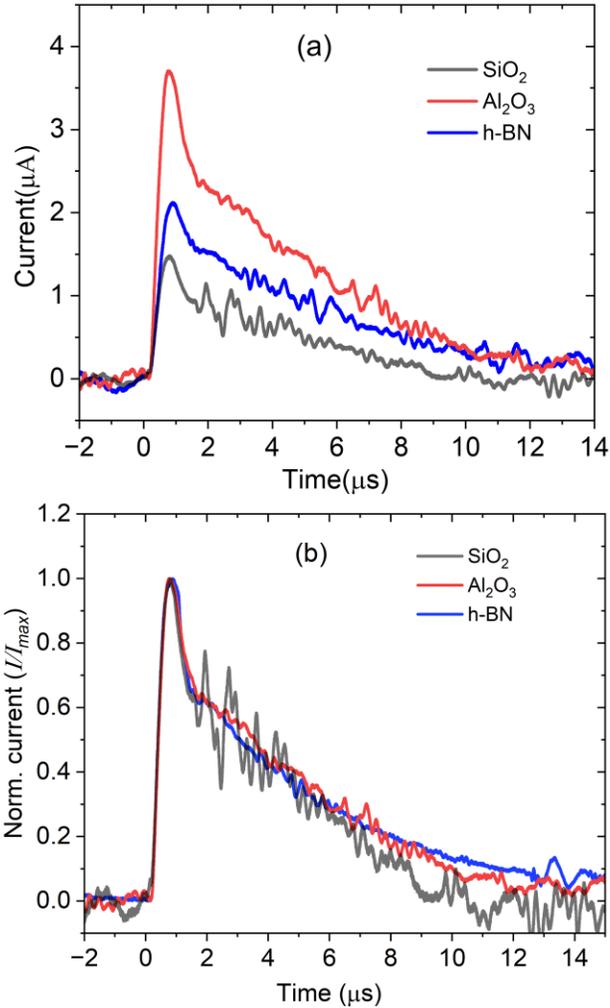

Figure 3. The photodetached electron current measured with TIA after first laser (355 nm) shot (a) and normalized values (b) for different samples.
The standard deviation of the curves is ±15%.

Table 3. The comparison of LSPD yield for single pulse obtained by ESVM and TIA measurements

| Sample | ESVM | TIA |
|---|---|---|
| $Al_2O_3$ | 4.89±0.83 x $10^{-9}$ | 3.57 ±1.24 x $10^{-9}$ |
| h-BN | 1.98±0.33 x $10^{-9}$ | 2.81±0.98 x $10^{-9}$ |
| $SiO_2$ | 4.51±0.9 x $10^{-10}$ | 1.185±0.41 x $10^{-10}$ |

Assuming no recharging of the surface by electrons during and after the surface exposure to the LSPD, the surface charge decay rate from the LSPD can be defined as[33]:

$$\frac{dN_e(t)}{dt} = -\Gamma_{ph}\sigma_{pd}N_e(t) \tag{3}$$

$$N_e(t) = N_e(0)\exp\left(\frac{-t}{\tau_{opt}}\right) + N_{e.sat} \tag{4}$$

where $N_e$ – surface density of electrons on the dielectric substrate, $N_{e.sat}$ – density of electrons after saturation of decay, $\Gamma_{ph}$ -photon flux per unit area, $\sigma_{pd}$ – photodetachment cross section for given laser photon energy. Then, the photodetachment decay of surface surplus electrons follow the exponential function with time constant defined by the cross section $\sigma_{pd}$ and photon flux $\Gamma_{ph}$:

$$\tau_{opt} = \frac{1}{\sigma\Gamma_{ph}} \tag{5}$$

The values of $\tau_{opt}$ can be obtained from the fitting results, photon flux obtained for assuming the ~9 ns of laser pulse width, and with assumption that the charge lost between laser pulses is negligible. Figure 4 demonstrates the exponential fitting of the surface electron charge decay curves driven by the LSPD for different materials and laser photon energy. The slight deviation of the fitting curves at larger pulse numbers might be related to the variation of the substrate temperature at high accumulated pulses, which may affect the decay process.

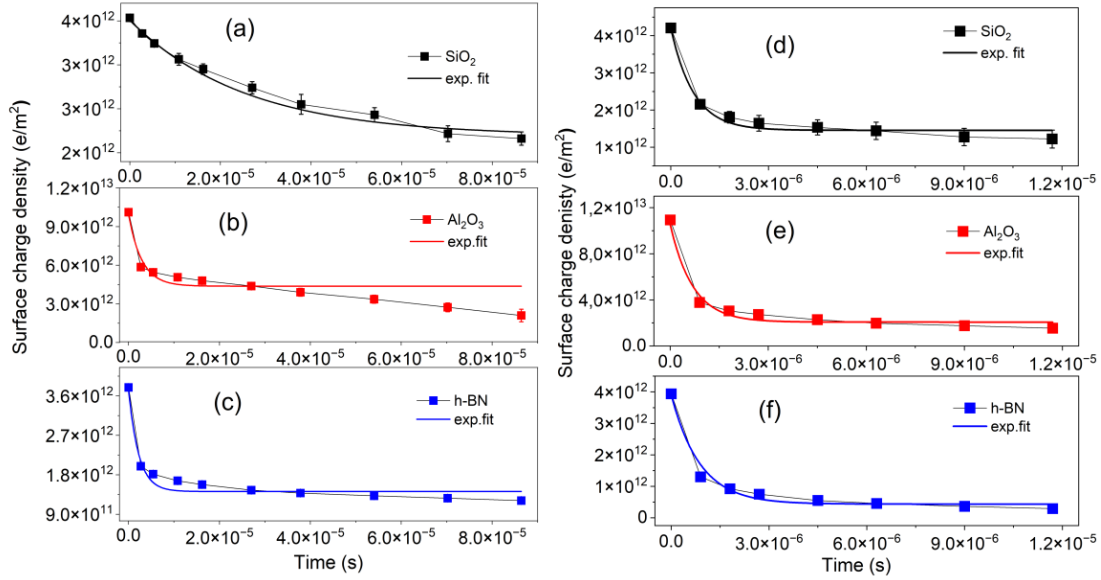

Figure 4. Fitting examples of LSPD induced electronic surface charge decay curves for different dielectric samples for 532 nm (a,b,c) and 355 nm (d,e,f) laser beams.

Following the above charge decay model, the average cross-section $\sigma_{pd}$ obtained for different laser fluences for three dielectric materials listed below in Table 4.

Table 4. The average LSPD cross-section $\sigma_{pd}$ for different samples

| Material | 355 nm (3.49 eV), $m^2$ | 532 nm (2.33 eV), $m^2$ |
| --- | --- | --- |
| SiO$_2$ | 2.99±0.72 x $10^{-23}$ | 0.69±0.2 x $10^{-24}$ |
| Al$_2$O$_3$ | 7.76±3.1 x $10^{-23}$ | 2.95±0.86 x $10^{-24}$ |
| h-BN | 7.65±3.3 x $10^{-23}$ | 3.16±1.06 x $10^{-24}$ |

For 532 nm laser ($h\nu$ =2.33 eV) the cross-section values are in the range of ~ $10^{-24}$ m$^2$ and for the 355 nm ($h\nu$ = 3.49 eV) are ~$10^{-23}$ m$^2$. The results show that the photodetachment cross section reaches the largest value for Al$_2$O$_3$ and h-BN, while it has the lowest value for the SiO$_2$ sample. The LSPD from different samples show that the photodetachment yield depends on the initial surface charge density on the surface. Highest yield shows the $Al_2O_3$ and h-BN, while the lowest for $SiO_2$, which is not in correlation with tabulated $\chi$ (Table 1).

To explain the nonlinear reduction of the surface charge decay rate and $\gamma_{pd}$ during the LSPD, we can assume a uniform distribution of charges on the substrate. The average values of $\sigma_e$ are 4 x $10^{12}$ e/m$^2$ for $SiO_2$ and h-BN, and 1.1 x $10^{13}$ e/m$^2$ for $Al_2O_3$. Considering the surface density of atoms on the crystal plane of the material is on the order of ~$10^{18}$ m$^2$, and the incident photon number density per laser pulse is ~ 2 - 5 x $10^{16}$ photons/m$^2$, the density of surplus electrons responsible for charging is much lower than the bulk electrons and photons incident on the surface during the LSPD. The surface potential decay measurements (Figure 2) show that the number of detached electrons depends on the sequence of laser shots, indicating a relatively high decay rate of surface negative potential during the initial pulses, followed by an exponential drop with subsequent shots. Further analysis of the LSPD induced current of detached electrons (detected by the TIA) for separate laser shots at 3.49 eV photon energy confirms this behavior. This can be explained by the fact that during the initial laser shots, the surface density of surplus electrons on the dielectric is relatively densely packaged, leading to a high probability of detachment. However, with subsequent laser shots, the surface electron density becomes lower than during the previous shots, resulting in a decrease in the probability of LSPD and lowering the $\gamma_{pd}$. From this, we can conclude that the photodetachment yield depends on the initial surface electron charge density at a fixed incident number of photons. A similar nonlinear reduction in surface charge decay rate is observed for the discharging of dielectric surfaces through natural charge decay[34] as well as through photon-excited[35] or thermally stimulated decay [36] processes.

Based on the tabulated $\chi$ values (Table 1), one would expect that $\gamma_{pd}$ should follow the same trend (lower electron affinity should result in a higher yield) at a given laser fluence and photon energy. The absence of the correlation between the $\chi$ and the $\gamma_{pd}$, may be explained by several factors. Firstly, the electron affinity $\chi$ of the material is a purely surface property and is highly sensitive to conditions including the presence of surface defects, impurities, and residuals from exposure to ambient atmosphere, plasma, and cleaning. Thus, effective $\chi$ may vary significantly for the same material depending on its physicochemical properties, surface functionalization, and the preparation methods[37,38,39]. Secondly, the hypothesis assuming $\chi$ as the "binding" energy of electrons and the threshold value for electron photodetachment overlooks the presence of electronic charge traps at the solid surface. These charge-trapping surface states are electronic energy states within the forbidden gap (with wide energy distribution) below the conduction band minimum, arising from surface defects, dangling bonds, surface adatoms[40]. Finally, the highest $\gamma_{pd}$ for $Al_2O_3$ is due to the the initial surface charge density which is almost ~3 times larger compared to that of h-BN and $SiO_2$. Consequently, the photodetachment yield is also higher at the same incident photon density (laser fluence). Although clean h-BN and $SiO_2$ have similar 'tabulated' electron affinities (Table 1), BN exhibits higher $\gamma_{pd}$ compared to quartz. This could be attributed to the fact that real samples of h-BN may have a lower χ due to surface impurities[41]. On the contrary, the $SiO_2$ samples may have high density deep intrinsic charge trap states (vacancies) due to the amorphous nature and high disorder in the surface crystal lattice structure[40,42]. The obtained photodetachment cross section values for all three samples are in the range of ~$10^{-23}$ - $10^{-24}$ $m^2$ for 3.49 eV and 2.33 eV photon energy respectively and demonstrated the similar trends as the photodetachment yield. Note, that typical values of the photodetachment cross section are in the range of ~$10^{-22}$ $m^2$ for the molecular ions of oxygen and hydrogen species[43]. It was also found to be 1.6 x $10^{-21}$ $m^2$ at 4.66 eV photon energy[27] for plasma grown dielectric nanoparticles, which is two orders of magnitude larger. This indicates that the obtained cross sections from our measurements are reasonable considering the lower photon energies.

In conclusion, LSPD induced surface charge decay rate from negatively charged dielectric substrates is nonlinear and decreases with subsequent laser shots. This indicates that for the given incident photon flux, the photodetachment yield can be highly affected by the initial density of surface charges. The photodetachment yields and cross sections for $Al_2O_3$, h-BN, and $SiO_2$ do not follow the tabulated electron affinity values; instead, they are defined by the surface physicochemical properties. The LSPD of electrons could alter the plasma properties by contributing to breakdown, sheath formation and ionization balance, particularly when the surface charge density is high due to the large capacitance of the insulating substrates (e.g., thin film) or applied high voltage. Additionally, there is a possibility that surface charge density could be controlled or even fully eliminated by properly adjusting flux of low-energy laser photons, with less damage to the surface. This could be useful for preventing arcing or charge-induced damage by surface charge control or minimization in for example, plasma processing and space applications. Regarding the LSPD diagnostics of nanoparticles in dusty plasmas, our results suggest that it is important to operate in the fully saturated regime to avoid underestimation of charge densities per particle. This is because not all surface electrons might be effectively removed by a single laser shot during LSPD, due to the low yield at relatively low surface charge density and the existence of deep electron traps. The obtained photodetachment yield and cross section values for common dielectric materials can be used as input data for modeling and simulations. Future work will focus on the study of LSPD directly in the plasma medium and on the understanding different additional photodetachment channels, such as laser-stimulated thermal emission and desorption of surface adatoms.

The authors acknowledge the US Department of Energy (DOE) for support of the work by the contract DE-AC02-09CH11466 and Mr. T. Bennet for providing technical assistance.

## Author Declarations

Conflict of Interest

The authors have no conflicts to disclose.

## Author Contributions

**Y. Ussenov**: Data curation (lead); Formal analysis (lead); Investigation (equal); Validation (equal); Writing – original draft (lead); Writing – review & editing (equal). **M. Shneider**: Conceptualization (equal); Investigation (equal); Formal analysis (equal); Writing – review & editing (equal); Supervision (equal). **S. Yatom**: Conceptualization (equal); Investigation (equal); Formal analysis (equal); Writing – review & editing (equal); Supervision (equal). **Y. Raitses**: Conceptualization (equal); Investigation (equal); Formal analysis (equal); Writing – review & editing (equal); Supervision (lead).

## Data Availability

The data that support the findings of this study are available from the corresponding author upon reasonable request.

# References

[1] F. X. Bronold, K. Rasek, H. Fehske, Electron microphysics at plasma–solid interfaces. *J. Appl. Phys.* **128 (18)** 180908 (2020)

[2] S. M. Rossnagel, J. J. Cuomo and W. D. Westwood, Handbook of Plasma Processing Technology: Fundamentals, Etching, Deposition, and Surface Interactions, Park Ridge: Noyes Publications, (1990)

[3] S. Nemschokmichal, R. Tschiersch, H. Höft et al., Impact of volume and surface processes on the pre-ionization of dielectric barrier discharges: advanced diagnostics and fluid modeling. *Eur. Phys. J. D*. **72**, 89 (2018)

[4] N. Marchack, L. Buzi, D. B. Farmer, H. Miyazoe, J. M. Papalia, H. Yan, G. Totir, S. U. Engelmann, Plasma processing for advanced microelectronics beyond CMOS. *J. Appl. Phys.* **130 (8)** 080901 (2021)

[5] G. S. Oehrlein and S. Hamaguchi, Foundations of low-temperature plasma enhanced materials synthesis and etching, *Plasma Sources Sci. Technol.* **27** 023001 (2018)

[6] S.-N. Hsiao, M. Sekine, K. Ishikawa, Y. Iijima, Y. Ohya, M. Hori, An approach to reduce surface charging with cryogenic plasma etching using hydrogen-fluoride contained gases, *Appl. Phys. Lett*. **123**, 212106 (2023)

[7] W. Zhu, S. Sridhar, L. Liu, E. Hernandez, V. M. Donnelly, D. J. Economou, Photo-assisted etching of silicon in chlorine- and bromine-containing plasmas, *J. Appl. Phys.* **115**, 203303 (2014)

[8] V. I. Demidov, S. V. Ratynskaia, K. Rypdal, Electric probes for plasmas: The link between theory and instrument, *Rev. Sci. Instrum*. **73**, 3409–3439 (2002)

[9] R. Brandenburg, Dielectric barrier discharges: progress on plasma sources and on the understanding of regimes and single filaments, *Plasma Sources Sci. Technol.* **26** 053001 (2017)

[10] Y. Raitses, A. Smirnov, D. Staack, N. J. Fisch, Measurements of secondary electron emission effects in the Hall thruster discharge, *Phys. Plasmas* **13**, 014502 (2006)

[11] S. T. Lai, Fundamentals of Spacecraft Charging: Spacecraft Interactions with Space Plasmas (Princeton, NJ: Princeton University Press) (2011)

[12] B. Horváth, M. Daksha, I. Korolov, A. Derzsi and J. Schulze, The role of electron induced secondary electron emission from $SiO_2$ surfaces in capacitively coupled radio frequency plasmas operated at low pressures, *Plasma Sources Sci. Technol.* **26** 124001 (2017)

[13] D.-Q. Wen, J. Krek, J. T. Gudmundsson, E. Kawamura, M. A. Lieberman, P. Zhang, J. P. Verboncoeur, Field reversal in low pressure, unmagnetized radio frequency capacitively coupled argon plasma discharges. *Appl. Phys. Lett.* **123** (26): 264102 (2023)

[14] A. Iqbal, B. Z Bentz, Y. Zhou, K. Youngman and P. Zhang, Pulsed photoemission induced plasma breakdown, *J. Phys. D: Appl. Phys.* **56** 505204 (2023)

[15] M. D. Campanell and M. V. Umansky, Strongly Emitting Surfaces Unable to Float below Plasma Potential, *Phys. Rev. Lett.* **116**, 085003 (2016)

[16] J. P. Sheehan, N. Hershkowitz, I. D. Kaganovich, H. Wang, Y. Raitses, E. V. Barnat, B. R. Weatherford, and D. Sydorenko, Kinetic Theory of Plasma Sheaths Surrounding Electron-Emitting Surfaces, *Phys. Rev. Lett.* **111**, 075002 (2013)

[17] A. V. Phelps and Z Lj Petrovic, Cold-cathode discharges and breakdown in argon: surface and gas phase production of secondary electrons, *Plasma Sources Sci. Technol.* **8** R21 (1999)

[18] A. Ottaviano, S. Banerjee, Y. Raitses; A rapid technique for the determination of secondary electron emission yield from complex surfaces. *J. Appl. Phys.*, **126** (22): 223301 (2019)

[19] A. Dunaevsky, Y. Raitses, N. J. Fisch, Secondary electron emission from dielectric materials of a Hall thruster with segmented electrodes, *Phys. Plasmas* **10**, 2574–2577 (2003)

[20] M. Bacal, Photodetachment diagnostic techniques for measuring negative ion densities and temperatures in plasmas, *Rev. Sci. Instrum.* **71** (11): 3981–4006 (2000)

[21] R. Tschiersch, S. Nemschokmichal and J. Meichsner, Influence of released surface electrons on the pre-ionization of helium barrier discharges: laser photodesorption experiment and 1D fluid simulation, *Plasma Sources Sci. Technol.* **26** 075006 (2017)

[22] O. Guaitella, I. Marinov, A. Rousseau, Role of charge photodesorption in self-synchronized breakdown of surface streamers in air at atmospheric pressure. *Appl. Phys. Lett.* **98** (7): 071502 (2011)

[23] Y.B. Golubovskii, V.A. Maiorov, J. Behnke and J.F. Behnke, Influence of interaction between charged particles and dielectric surface over a homogeneous barrier discharge in nitrogen, *J. Phys. D: Appl. Phys.* **61** 35751(2002)

[24] P. F. Ambrico, M. Ambrico, L. Schiavulli, T. Ligonzo, V. Augelli, Charge trapping induced by plasma in alumina electrode surface investigated by thermoluminescence and optically stimulated luminescence. *Appl. Phys. Lett.*, **94** (5): 051501 (2009)

[25] M. Li, C. Li, H. Zhan, J. Xu, X. Wang, Effect of surface charge trapping on dielectric barrier discharge. *Appl. Phys. Lett.*, **92** (3): 031503 (2008)

[26] E. Stoffels, W. W. Stoffels, G. M. W. Kroesen, F. J. de Hoog, Dust formation and charging in an Ar/SiH4 radio-frequency discharge, *J. Vac. Sci. Technol. A* **14**, 556–561 (1996)

[27] T. J. A. Staps, T. J. M. A. Donders, B. Platier and J. Beckers, In-situ measurement of dust charge density in nanodusty plasma, *J. Phys. D: Appl. Phys.* **55** 08LT01 (2022)

[28] M. Shneider, Y. Raitses, S. Yatom, Schottky effect on the wavelength threshold for the photo-detachment from charged metallic nanoparticles, *J. Phys. D: Appl. Phys.* **56** (29), 29LT01 (2023); *J. Phys. D: Appl. Phys.* **56** 439501 (2023)

[29] F. X. Bronold and H. Fehske, Absorption of an Electron by a Dielectric Wall, *Phys. Rev. Lett.* **115**, 225001 (2015)

[30] R. L. Heinisch, F. X. Bronold, and H. Fehske, Electron surface layer at the interface of a plasma and a dielectric wall, *Phys. Rev. B* **85**, 075323 (2012)

[31] I. Shlyakhov, J. Chai, M. Yang, S. Wang, V. V. Afanas'ev, M. Houssa, A. Stesmans, Energy band alignment of a monolayer $MoS_2$ with $SiO_2$ and $Al_2O_3$ Insulators from Internal Photoemission, *Phys. Status Solidi A*, **216** 1800616 (2019)

[32] T. Knobloch, Y.Y. Illarionov, F. Ducry et al., The performance limits of hexagonal boron nitride as an insulator for scaled CMOS devices based on two-dimensional materials. *Nat. Electron.* 4, 98–108 (2021).

[33] H. Amjadi, The mechanism of voltage decay in corona-charged layers of silicon dioxide during UV irradiation, *IEEE Transactions on Dielectrics and Electrical Insulation*, **7**(2) 222 – 228 (2000)

[34] J. Kindersberger, C. Lederle, Surface charge decay on insulators in air and sulfurhexafluorid - part I: simulation, *IEEE Transactions on Dielectrics and Electrical Insulation,* **15** (4), 941-948 (2008)

[35] S. Lin, L. Xu, L. Zhu, X. Chen, and Z. L. Wang, Electron Transfer in Nanoscale Contact Electrification: Photon Excitation Effect, *Adv. Mater.* **31** 1901418 (2019)

[36] S. Lin, L. Xu, C. Xu, X. Chen, A. C. Wang, B. Zhang, P. Lin, Ya Yang, H. Zhao, Z. Lin Wang, Electron Transfer in Nanoscale Contact Electrification: Effect of Temperature in the Metal–Dielectric Case, *Adv. Mater.* **31**, 1808197 (2019).

[37] L. Giordano, P. V. Sushko, G. Pacchioni, and A. L. Shluger, Electron Trapping at Point Defects on Hydroxylated Silica Surfaces, *Phys. Rev. Lett.* **99**, 136801 (2007)

[38] J. A. Sedlacek, E. Kim, S. T. Rittenhouse, P. F. Weck, H. R. Sadeghpour, and J. P. Shaffer, Electric Field Cancellation on Quartz by Rb Adsorbate-Induced Negative Electron Affinity, *Phys. Rev. Lett.* **116**, 133201 (2016)

[39] F. Buonocore, A. Capasso, M. Celino, N. Lisi, and O. Pulci, Tuning the Electronic Properties of Graphane via Hydroxylation: An Ab Initio Study, *J. Phys.Chem.C* ,**125**,16316−16323 (2021)

[40] Jack Strand et al., Intrinsic charge trapping in amorphous oxide films: status and challenges, *J. Phys.: Condens. Matter* **30** 233001 (2018)

[41] K. P. Loh, I. Sakaguchi, M. N. Gamo, S. Tagawa, T. Sugino, T. Ando; Surface conditioning of chemical vapor deposited hexagonal boron nitride film for negative electron affinity. *Appl. Phys. Lett.* **74** (1): 28–30 (1999)

[42] A. Kimmel et. al., Positive and Negative Oxygen Vacancies in Amorphous Silica, *ECS Trans.* **19** 3 (2009)

[43] L. C. Lee, G. P. Smith, Photodissociation and photodetachment of molecular negative ions. VI. Ions in $O_2/CH_4/H_2O$ mixtures from 3500 to 8600 Å. *J. Chem. Phys.* **15**, 70 (4): 1727–1735 (1979)